\documentstyle[preprint,aps,prb,eqsecnum,epsf]{revtex}
\begin{document}
\draft

\title{Metamagnets in uniform and random fields }

\author{Serge Galam,}
\address{Laboratoire des Milieux Desordonn\'{e}s et H\'{e}t\'{e}rog\`{e}nes\\
Universit\'{e} de Paris 6\\ Case 86, T13\\ 4 Place Jussieu\\ Cedex 05,
Paris, France}

\author{Carlos S. O. Yokoi, and Silvio R. Salinas}
\address{Instituto de F\'{\i}sica\\
Universidade de S\~{a}o Paulo\\ Caixa Postal 66318\\ 05315-970,
S\~{a}o Paulo, SP, Brazil}

\date{\today}
\maketitle

\begin{abstract}
We study a two-sublattice Ising metamagnet with nearest and
next-nearest-neighbor interactions, in both uniform and random
fields. Using a mean-field approximation, we show that the qualitative
features of the phase diagrams are significantly dependent on the
distribution of the random fields. In particular, for a Gaussian
distribution of random fields, the behavior of the model is
qualitatively similar to a dilute Ising metamagnet in a uniform field.


\end{abstract}


\section{Introduction}

The random-field Ising model has been a considerable source of
research over the last twenty years \cite{shapir92,gofman96,newman96}.
Systems with quenched random fields are experimentally realized in
antiferromagnets with bond mixing or site dilution
\cite{fishman79,cardy84}.  A large variety of these systems have been
subjected to detailed experimental studies \cite{belanger92}.

Although most theoretical problems associated with the ferromagnetic
Ising model in a random field (as the lower critical dimension, the
pinning effects, and the existence of a static phase transition) have
been solved, some questions are still open. In particular, there is
still room to investigate the existence of a tricritical point
\cite{maritan91,swift94,swift97} and the exact relation to the
dilute antiferromagnet in a uniform field. Depending on the choice of
the random-field distribution, the mean-field approximation gives rise
to a tricritical point (which is present for a symmetric double-delta
distribution \cite{aharony78}, but does not occur in the case of a
Gaussian form \cite{schneider77}). On the basis of the central limit
theorem, some hand-wave arguments can be used to support the physical
relevance of the Gaussian distribution (the tricritical point produced
by the double-delta functions being a mere artifact of the mean-field
approximation).

The proof of the equivalence between an Ising ferromagnet in a random
field and a dilute antiferromagnet in a uniform field is based on
renormalization-group arguments that can be applied for weak fields
\cite{fishman79,cardy84}. In the mean-field approximation
\cite{galam85a}(or in the equivalent and exactly soluble model with
infinite-range interactions \cite{perez86,matos92}), it is possible to
establish a complete mapping between the parameters of the Ising
ferromagnet in a random field and the dilute Ising antiferromagnet or
metamagnet in a uniform field. In particular, it is known that the
random fields should be associated with a symmetric double-delta
distribution for arbitrary dilution \cite {galam85a,perez86,matos92},
including the pure case where there is no dilution! This peculiar
result suggests that, instead of describing the random fields
generated by dilution, the mean-field approximation is just referring
to the two-sublattice structure of the antiferromagnet (which is
reflected in the symmetric double-delta distribution of the random
fields).  It should be mentioned that the mean-field approximation for
the dilute Ising metamagnet in a uniform field suffers from other
difficulties when confronted with Monte Carlo calculations
\cite{diep87,diep88,galam89} and experimental results
\cite{kushauer95,mattsson96}. Whereas numerical simulations and
experiments indicate that the first-order transition is destroyed when
the dilution is increased, no such effect is predicted in a mean-field
calculation.

In this paper we use a mean-field approximation to consider an Ising
metamagnet with nearest and next-nearest-neighbor interactions, in a
uniform field and a random field. This model is equivalent to a dilute
Ising metamagnet in a field for an appropriate choice of the random
field distribution. Since the exact mapping of the dilution to the
random fields is unknown, only a qualitative comparison can be
made. We consider double-delta and Gaussian random-field
distributions. The behavior of the model and the phase diagrams depend
very much on these random-field distributions. The Gaussian form seems
to be more appropriate for a description of the diluted system.


\section{Definition of the model}

We consider a regular lattice of $N$ sites, with Ising spins $S_i=\pm
1$ at each site, that can be divided into two equivalent
interpenetrating sublattices, A and B. The $z$ nearest neighbor
(nn) spins of a given spin are on the other sublattice, while the
$z^{\prime }$ next-nearest neighbor (nnn) spins are all on the same
sublattice. The Hamiltonian of the system is given by

\begin{equation}
{\cal H}=J\sum_{{\rm nn}}S_iS_j-J^{\prime }\sum_{{\rm nnn}
}S_iS_j-\sum_i(H+H_i)S_i,  \label{HAMILTONIAN}
\end{equation}
where $J$ is the nn exchange parameter, the sum $\sum_{{\rm nn}}$ is
over all pairs of nn spins, $J^{\prime }$ is the nnn exchange
parameter, the sum $\sum_{{\rm nnn}}$ is over all nnn spins, $H$ is
the strength of the external uniform magnetic field, and $H_i$ is the
strength of the local random field.  We assume that the nn
interactions are antiferromagnetic ($J>0$), the nnn interactions are
ferromagnetic ($J^{\prime }\ge 0$), and the local random fields $H_i$
are uncorrelated. Even though it is possible to consider
sublattice-dependent probability distributions, in this paper we use
the same probability distribution at every site.


\section{Mean-field equations}

We derive the mean-field equations from Bogoliubov's variational principle 
\cite{callen85}, 

\begin{equation}
\langle F \rangle_{\rm av} \le \langle
F_{\rm t} \rangle_{\rm av} + \langle \langle \cal{H} -
\cal{H}_{\rm t} \rangle_{\rm t} \rangle_{\rm av},
\end{equation}
where $\langle \cdots \rangle_{\rm av}$ denotes averaging over
the random-field distribution and $\langle \cdots
\rangle_{\rm t}$ the thermal averaging with respect to the trial
Hamiltonian ${\cal H}_{\rm t}$. Choosing the non-interacting
trial Hamiltonian

\begin{equation}
{\cal H}_{{\rm t}}=-\sum_i(H+H_i)S_i-\eta _{{\rm A}}\sum_{i\in {\rm A}
}S_i-\eta _{{\rm B}}\sum_{i\in {\rm B}}S_i,  \label{TRIALHAMILTONIAN}
\end{equation}
where $\eta _{{\rm A}}$ and $\eta _{{\rm B}}$ are the variational
parameters, we obtain

\begin{eqnarray}
\langle F \rangle_{{\rm av}} &\le& -\frac{N}{2 \beta} \langle \ln 2 \cosh
\beta (H+H_i+\eta_{{\rm A}}) \rangle_{{\rm av}}  \nonumber \\
& & - \frac{N}{2 \beta} \langle \ln 2 \cosh \beta (H+H_i+\eta_{{\rm B}})
\rangle_{{\rm av}} + \frac{J N z}{2} m_{{\rm A}} m_{{\rm B}}  \nonumber \\
& & - \frac{J^{\prime}N z^{\prime}}{4} (m_{{\rm A}}^2 + m_{{\rm B}}^2) + 
\frac{N}{2} \eta_{{\rm A}} m_{{\rm A}} + \frac{N}{2} \eta_{{\rm B}} m_{{\rm B
}},  \label{INEQUALITY}
\end{eqnarray}
with

\begin{mathletters}
\begin{eqnarray}
m_{{\rm A}} &=&\langle \tanh \beta (H+H_i+\eta _{{\rm A}})\rangle _{{\rm av}
},  \label{MA} \\
m_{{\rm B}} &=&\langle \tanh \beta (H+H_i+\eta _{{\rm B}})\rangle _{{\rm av}
}.  \label{MB}
\end{eqnarray}
The condition that the right-hand side of Eq. (\ref{INEQUALITY}) is
stationary determines the variational parameters,

\end{mathletters}
\begin{mathletters}
\begin{eqnarray}
\eta _{{\rm A}} &=&-Jzm_{{\rm B}}+J^{\prime }z^{\prime }m_{{\rm A}},
\label{ETAA} \\
\eta _{{\rm B}} &=&-Jzm_{{\rm A}}+J^{\prime }z^{\prime }m_{{\rm B}}.
\label{ETAB}
\end{eqnarray}
Inserting Eqs. (\ref{ETAA})-(\ref{ETAB}) into Eqs. (\ref{MA})-(\ref{MB}), we
arrive at the mean-field equations,

\end{mathletters}
\begin{mathletters}
\begin{eqnarray}
m_{{\rm A}} &=&\langle \tanh \beta (H+H_i-Jzm_{{\rm B}}+J^{\prime }z^{\prime
}m_{{\rm A}})\rangle _{{\rm av}},  \label{MFMA} \\
m_{{\rm B}} &=&\langle \tanh \beta (H+H_i-Jzm_{{\rm A}}+J^{\prime }z^{\prime
}m_{{\rm B}})\rangle _{{\rm av}}.  \label{MFMB}
\end{eqnarray}
The right-hand side of Eq. (\ref{INEQUALITY}) at the stationary point gives
the mean-field free energy per spin,

\end{mathletters}
\begin{eqnarray}
f &=& - \frac{1}{2 \beta} \langle \ln 2 \cosh \beta (H+H_i - J z m_{{\rm B}}
+ J^{\prime}z^{\prime}m_{{\rm A}} ) \rangle_{{\rm av}}  \nonumber \\
& & \mbox{} - \frac{1}{2 \beta} \langle \ln 2 \cosh \beta (H+H_i - J z m_{
{\rm A}} + J^{\prime}z^{\prime}m_{{\rm B}}) \rangle_{{\rm av}}  \nonumber \\
& & \mbox{} - \frac{J z}{2} m_{{\rm A}} m_{{\rm B}} + \frac{
J^{\prime}z^{\prime}}{4} (m_{{\rm A}}^2 + m_{{\rm B}}^2).
\label{MFFREEENERGY}
\end{eqnarray}


\section{Landau expansion}

\begin{figure}
\centerline{\setlength{\epsfysize}{6cm}
\epsfbox{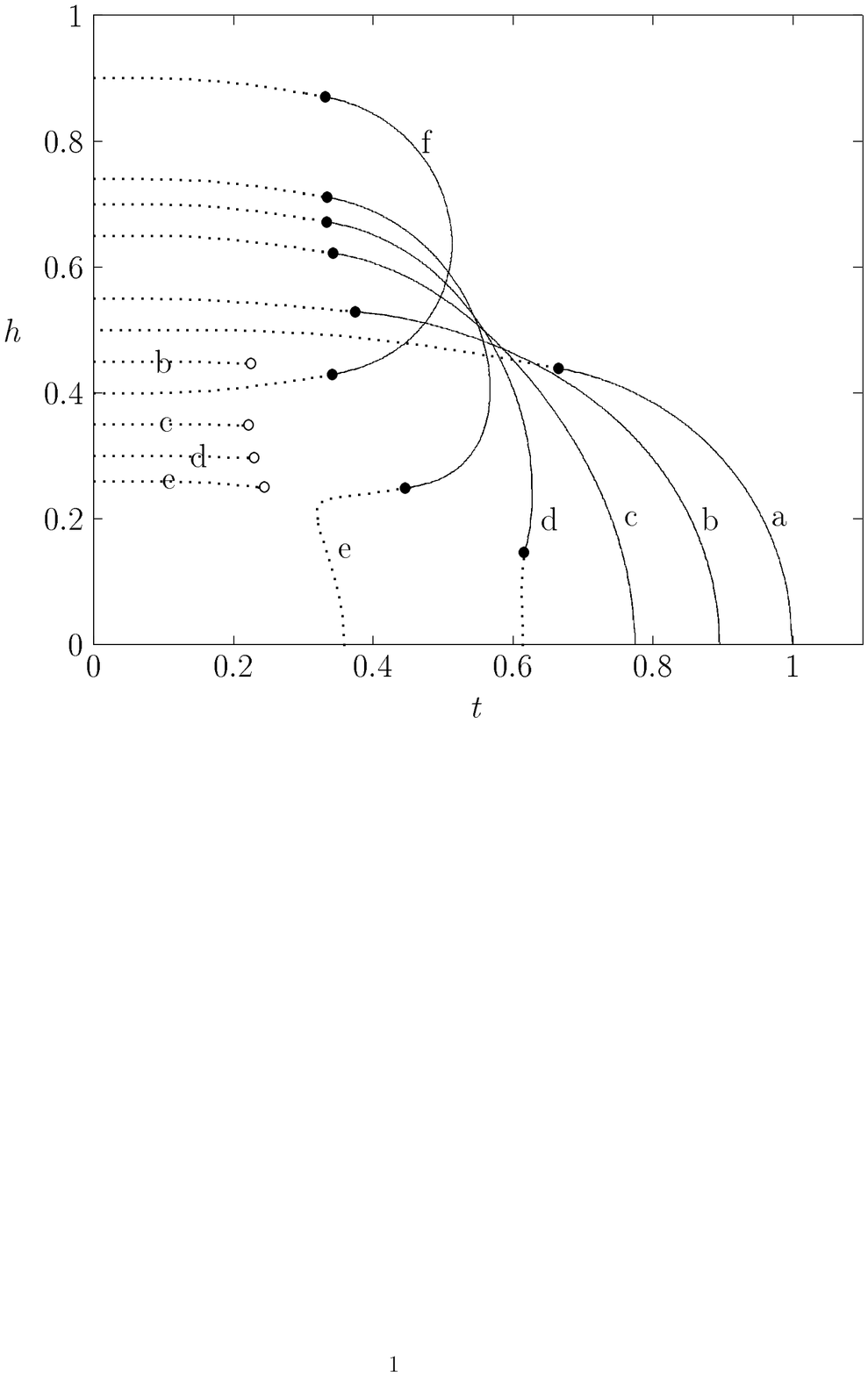}}
\caption{Phase diagrams in $h-t$ plane in the case of a double-delta
distribution for (a) $\sigma =0$, (b) $\sigma =0.3$, (c) $\sigma =0.4$, (d) $
\sigma =0.45$, (e) $\sigma =0.49$ and (f) $\sigma =0.65$. The solid lines
represent continuous transitions. The dashed lines are first-order
transitions. The filled circles are tricritical points, and the empty
circles are critical points.}
\label{DHXT}
\end{figure}

In this Section we develop the Landau expansion along the same steps used
for the pure case \cite{kincaid75}. It is convenient to introduce the
reduced quantities

\begin{equation}
t = \frac{1}{\beta(Jz+J^{\prime}z^{\prime})}, \quad h = \frac{H}{
Jz+J^{\prime}z^{\prime}} , \quad h_i = \frac{H_i}{Jz+J^{\prime}z^{\prime}},
\end{equation}
and the parameters

\begin{equation}
\epsilon =\frac{J^{\prime }z^{\prime }}{Jz}\ge 0,\quad \gamma =\frac{
-Jz+J^{\prime }z^{\prime }}{Jz+J^{\prime }z^{\prime }}=\frac{\epsilon -1}{
\epsilon +1}.
\end{equation}
In terms of the uniform and staggered magnetizations,

\begin{equation}
M=\frac{m_{{\rm A}}+m_{{\rm B}}}2,\quad m_{{\rm s}}=\frac{m_{{\rm A}}-m_{
{\rm B}}}2,
\end{equation}
the mean-field equations (\ref{MFMA}) and (\ref{MFMB}) can be written as

\begin{equation}
M=\frac 12\left[ \langle \tanh \frac 1t(h+h_i+\gamma M+m_{{\rm s}})\rangle _{
{\rm av}}+\langle \tanh \frac 1t(h+h_i+\gamma M-m_{{\rm s}})\rangle _{{\rm av
}}\right] ,  \label{BIGM}
\end{equation}
and 
\begin{equation}
m_{{\rm s}}=\frac 12\left[ \langle \tanh \frac 1t(h+h_i+\gamma M+m_{{\rm s}
})\rangle _{{\rm av}}-\langle \tanh \frac 1t(h+h_i+\gamma M-m_{{\rm s}
})\rangle _{{\rm av}}\right] .  \label{MS}
\end{equation}
Also, the free energy per spin, given by Eq. (\ref{MFFREEENERGY}), may be
written in the form

\begin{eqnarray}
f & = & -\frac{t}{2} \langle \ln 2 \cosh \frac{1}{t} (h+h_i + \gamma M + m_{
{\rm s}} ) \rangle_{{\rm av}}  \nonumber \\
& & \mbox{} - \frac{t}{2} \langle \ln 2 \cosh \frac{1}{t} (h+h_i+ \gamma M -
m_{{\rm s}}) \rangle_{{\rm av}} - \frac{\gamma}{2} M^2 + \frac{1}{2} m_{{\rm s}
}^2.  \label{FREEENERGY}
\end{eqnarray}
Let us now write the uniform magnetization as $M=M_0+m$, where $M_0$ is the
paramagnetic solution, given by equation

\begin{equation}
M_0=\langle \tanh \frac 1t(h+h_i+\gamma M_0)\rangle _{{\rm av}}.
\end{equation}
The expansion of the right-hand side of Eqs. (\ref{BIGM}) and (\ref{MS}) in
powers of $(\gamma m\pm m_{{\rm s}})$ gives the expressions

\begin{mathletters}
\begin{eqnarray}
m & = & \frac{1}{2} \sum_{n=1}^\infty A_n [ ( \gamma m + m_{{\rm s}})^n +
(\gamma m - m_{{\rm s}})^n],  \label{MSUM} \\
m_{{\rm s}} & = & \frac{1}{2} \sum_{n=1}^\infty A_n [ ( \gamma m + m_{{\rm s}
})^n - (\gamma m - m_{{\rm s}})^n],  \label{MSSUM}
\end{eqnarray}
where the coefficients $A_n$ are given by

\end{mathletters}
\begin{mathletters}
\begin{eqnarray}
A_1 & = & -\frac{1}{t}(T_2-1), \\
A_2 & = & \frac{1}{t^2}(T_3-T_1), \\
A_3 & = & -\frac{1}{3 t^3}(3 T_4-4T_2+1), \\
A_4 & = & \frac{1}{3 t^4}(3 T_5-5T_3+2T_1), \\
A_5 & = & -\frac{1}{15 t^5}(15T_6-30T_4+17T_2-2).
\end{eqnarray}
with

\end{mathletters}
\begin{equation}
T_k = \langle \tanh^k \frac{1}{t} (h + h_i + \gamma M_0) \rangle_{{\rm av}}.
\end{equation}
We now determine $m$ in terms of $m_{{\rm s}}$ in the form

\begin{equation}
m=B_1m_{{\rm s}}^2+B_2m_{{\rm s}}^4+B_3m_{{\rm s}}^6+\ldots .
\label{MEXPANDED}
\end{equation}
Inserting this expansion into Eq. (\ref{MSUM}), and equating the
coefficients of same degree in $m_{{\rm s}}$, we find the coefficients $B_n$
in terms of $A_n$. Finally, substituting $m$, given by Eq. (\ref{MEXPANDED}
), into Eq.(\ref{MSSUM}) we obtain the expansion

\begin{equation}
a m_{{\rm s}} + b m_{{\rm s}}^3 + c m_{{\rm s}}^5 + \cdots = 0,
\end{equation}
where

\begin{mathletters}
\begin{eqnarray}
a &=&1-A_1, \\
b &=&\frac{2\gamma A_2^2}{\gamma A_1-1}-A_3, \\
c &=&\frac{2\gamma ^3A_2^4}{(\gamma A_1-1)^3}-\frac{9\gamma ^2A_2^2A_3}{
(\gamma A_1-1)^2}+\frac{6\gamma A_2A_4}{\gamma A_1-1}-A_5.
\end{eqnarray}
The second order transition is found at $a=0$ with $b>0$. The tricritical
point occurs for $a=b=0$ with $c>0$.

In the absence of random fields the model exhibits a tricritical point \cite
{kincaid75} in the $h-t$ phase diagram for $\epsilon >3/5$. In the numerical
calculations of the next sections, we just consider the case $\epsilon =1$,
which is typical for the range of values $\epsilon >3/5$.


\section{Phase diagrams for the double-delta distribution}

In this Section we study the phase diagrams for the case of a double-delta
distribution,

\end{mathletters}
\begin{equation}
P(h_i)=\frac 12[\delta (h_i-\sigma )+\delta (h_i+\sigma )].
\end{equation}
Fig. \ref{DHXT} shows the phase diagrams in the $h-t$ plane for various
values of the randomness $\sigma $.

The case of no randomness ($\sigma =0$) corresponds to the pure Ising
metamagnet \cite{kincaid75}. The phase diagram comprises a
metamagnetic phase ($m_{\rm s} \ne 0$) at low fields and a
paramagnetic phase ($m_{\rm s}=0$) at high fields. The transitions
between these phases are first-order for low temperatures and
second-order for high temperatures, being separated by a tricritical
point \cite{kincaid75} at $t=2/3$, as shown in Fig. \ref {DHXT}(a).

For $\sigma >(2/3)\tanh {}^{-1}1/\sqrt{3}=0.438\ldots $, there is a
second tricritical point at lower fields, as illustrated by
Figs. \ref{DHXT}(d)--(e). Also, for the randomness in the interval
$0<\sigma <0.5$, there is a first-order transition line inside the
metamagnetic phase at low temperatures. Through this transition line
the staggered magnetization decreases discontinuously as the field is
increased. This internal first-order transition line ends at a
critical point. Finally, for $\sigma >0.5$, the internal and lower
first-order transition lines merge into a single first-order
transition line ending at the tricritical point, as illustrated in
Fig. \ref{DHXT}(f).

For the particular case of the double-delta distribution and $\epsilon
=1$ or $\gamma =0$ that we are considering, the phase diagrams in the
$\sigma -t$ plane are exactly the same as in the $h-t$ plane. This
comes from the invariance of Eqs. (\ref{MS}) and (\ref{FREEENERGY}),
for the staggered magnetization and the free energy, respectively,
under the interchange between $h$ and $\sigma $ (and from the
independence of the free energy on the uniform magnetization
$M$). Therefore, Fig. \ref{DHXT} also represents the phase diagrams in
the $\sigma -t$ plane if we interchange $h$ and $\sigma $ throughout
this figure and in its caption. The phase diagram comprises a
metamagnetic phase ($m_{{\rm s}}\ne 0$) for small $\sigma $ and a
paramagnetic phase ($m_{{\rm s}}=0$) for high values of $\sigma $. In
particular, the phase diagram for $h=0$ is equivalent (after flipping
all the spins on a sublattice) to the diagram of a ferromagnetic Ising
model in a double-delta random field \cite{aharony78}.


\section{Phase diagrams for the Gaussian distribution}

\begin{figure}
\centerline{\setlength{\epsfysize}{6cm}
\epsfbox{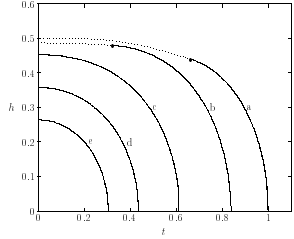}}
\caption{Phase diagrams in $h-t$ plane in the case of a Gaussian
distribution for (a) $\sigma = 0$, (b) $\sigma = 0.4$, (c) $\sigma = 0.6$,
(d) $\sigma = 0.7$ and (e) $\sigma = 0.75$. The solid lines represent
continuous transitions. The dashed lines are first-order transitions. The
filled circles are tricritical points.}
\label{GHXT}
\end{figure}

Now we study the phase diagrams for the Gaussian distribution,

\begin{equation}
P(h_i)=\frac 1{\sqrt{2\pi }\sigma }\exp (-\frac{h_i^2}{2\sigma ^2}).
\end{equation}
In Fig. \ref{GHXT}, we show the $h-t$ plane for various values of the
randomness $\sigma $. Again, the case of no randomness ($\sigma =0$)
corresponds to the pure Ising metamagnet \cite{kincaid75}, with a
first-order separated from a second-order transition line by a tricritical
point at $t=2/3$. The tricritical temperature decreases as the randomness is
increased until $\sigma =0.5$, when the transition between the metamagnetic
and paramagnetic phases becomes everywhere of second-order. The similarity
of these phase diagrams as a function of $\sigma $ with those of a dilute
metamagnet as a function of dilution \cite{diep87,diep88,galam89} is quite
striking. It suggests that a Gaussian random field gives, at least
qualitatively, a good description of the random fields generated by dilution
in a metamagnet.

\begin{figure}
\centerline{\setlength{\epsfysize}{6cm}
\epsfbox{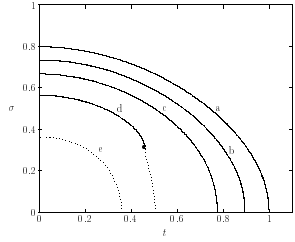}}
\caption{Phase diagrams in $\sigma-t$ plane in the case of a Gaussian
distribution for (a) $h = 0$, (b) $h = 0.3$, (c) $h = 0.4$, (d) $h = 0.47$
and (e) $h = 0.49$. The solid lines represent continuous transitions. The
dashed lines are first-order transitions. The filled circle is a tricritical
point.}
\label{GSIGMAXT}
\end{figure}

In Fig. \ref{GSIGMAXT}, we show the phase diagram in the $\sigma -t$ plane
for various values of the uniform field $h$. The case $h=0$ is equivalent
(after flipping all the spins on a sublattice) to the ferromagnetic Ising
model in a Gaussian random field \cite{schneider77}. The transition line is
of second-order for all temperatures and it crosses the $\sigma $ axis at $
\sigma =\sqrt{2/\pi }=0.79\ldots $. For large $\sigma$  the transition
at low fields becomes first-order and a tricritical point separates the
first-order and the second-order lines. For still larger randomness, the
transition becomes first-order always.


\section{Conclusions}

We have used the mean-field approximation to show that the phase diagrams of
an Ising metamagnet in the presence of a uniform and of random fields are
strongly dependent on the form of the distribution of probabilities of the
random fields. In particular, if the model exhibits a first-order transition
in zero random field, then a double-delta distribution never destroys this
first-order transition, in contradistinction to the case of a Gaussian
distribution. In this respect, there is a striking similarity in the
qualitative behavior of the metamagnet in a Gaussian random field and a
dilute metamagnet. This suggests that, by keeping the two-sublattice
structure and choosing an appropriate random field distribution, we can give
a better description of the dilute metamagnet than the previous mean-field
studies that map dilute Ising metamagnets in a uniform field into Ising
ferromagnets in a double-delta distribution of random fields.


\section*{\bf Acknowledgments}

This work was supported by the program USP/COFECUB (Grant No.
94.1.19622.1.0--UC--12/94). C.S.O.Y. and S.R.S. acknowledge partial
financial support from Conselho Nacional de Desenvolvimento
Cient\'{\i}fico e Tecnol\'ogico (CNPq) and Funda\c{c}\~ao de Amparo \`a
Pesquisa do Estado de S\~ao Paulo (FAPESP).



\newpage


\newpage


\newpage


\end{document}